\documentclass[preprint,prd,showpacs,showkeys]{revtex4-1}
\usepackage{amsmath}
\usepackage{amssymb}
\usepackage{amsfonts}
\usepackage{graphicx}
\usepackage{appendix}

\begin{document}

\title{Higgs-Photon-Photon Amplitude in the Unitary Gauge}
\author{Er-Cheng Tsai}
\email{ectsai@ntu.edu.tw}
\affiliation{Physics Department, National Taiwan University, Taipei, Taiwan}

\begin{abstract}
The physical process of Higgs decaying to two photons is the most important
mode in the discovery of the Higgs particle. Theoretical calculation of
$H\rightarrow\gamma\gamma$ amplitude therefore provides us with an avenue to
explore the underlying theory. Yet some confusion has arisen over the gauge
invariant property of $H\rightarrow\gamma\gamma$ in the standard model. In
this paper, we show that if dimensional regularization is adopted, the
on-shell 1-loop $H\rightarrow\gamma\gamma$ amplitude calculated in the
standard model is gauge invariant. In particular, we prove that the physical
amplitude calculated in the unitary gauge is the same as that calculated in
the Feynman gauge.
\end{abstract}

\pacs{11.10.Kk, 11.15.-q, 11.15.Bt, 12.15.Lk}
\keywords{quantum field; unitary gauge; dimensional regularization; electroweak }
\maketitle
\affiliation{Physics Department, National Taiwan University, Taipei, Taiwan}

\section{Introduction}

The gauge symmetry of quantum gauge field theory can be established
perturbatively \cite{WTI}. In the diagrammatic proof, we only need to assume
the legitimacy of shifting loop momenta to guarantee gauge invariance of
Feynman integrands for the physical amplitudes. In order to shift or reflect
loop variables for a Feynman integrand that does not give convergent integral,
a gauge invariant regularization scheme must be utilized. Dimensional
regularization \cite{HV} is such a scheme that can be implemented readily
because it does not introduce additional terms in the Lagrangian and hence
does not alter the forms of Feynman integrands when the space-time dimension
$n\neq4$.

The physical process of Higgs \cite{HIGGS} decaying to two photons is the most
important mode in the discovery of the Higgs particle \cite{FoundH}. But the
experimental result of $H\rightarrow\gamma\gamma$ amplitude did not seem to
agree with the theoretical 1-loop result obtained from the standard model. For
this T. T. Wu and S. L. Wu \cite{TTW}\ argued that the result obtained in the
unitary gauge \cite{UG} with certain assignments of loop variables differs
from that obtained from other covariant $R_{\xi}$\ gauges and may be used to
explain the discrepancy.

The unitary gauge can be considered as the limit $\xi\rightarrow\infty$ of the
$R_{\xi}$ gauge. Under dimensional regularization, a Feynman integrand can be
set to its value in the limit $\xi\rightarrow\infty$ before loop integrations.
This is in general more convenient than taking the limit $\xi\rightarrow
\infty$ after loop integrations as we can ignore the diagrams with the
unphysical field lines \cite{GIU}. In this paper, we show that the physical
$H\rightarrow\gamma\gamma$ amplitude calculated in the Feynman gauge is the
same as that in the unitary gauge provided dimensional regularization is
adopted. The unitary gauge amplitude remains gauge invariant.

\section{Integrands for 1-Loop $H\rightarrow\gamma\gamma$ Diagrams}

We now proceed to write out the integrands for all the diagrams contributing
to the 1-loop $H\rightarrow\gamma\gamma$ amplitude for the electroweak theory
defined in the Appendices. For the unitary gauge, only two diagrams need to be
considered (Fig. $\ref{hgg-w})$: one with 3 internal vector meson lines, the
other with 2 internal vector meson lines. For the Feynman gauge ($R_{1}$
gauge), there are 11 additional diagrams (Fig. $\ref{hgg-fiG})$ with some
internal lines being ghost or non-physical field lines. Under dimensional
regularization, we are allowed to the shift loop variable $\ell$ by any
amount. We shall also make use of the refection symmetry $\ell_{\sigma
}\rightarrow-\ell_{\sigma}$ for any component of the loop momentum to ignore a
Feynman integrand that is odd under $\ell\rightarrow-\ell,$%
\begin{equation}
\ell_{\mu}f\left(  \ell^{2}\right)  \rightarrow0,\;\ell_{\mu}\ell_{\nu}%
\ell_{\rho}f\left(  \ell^{2}\right)  \rightarrow0,\;\ell_{\mu}\ell_{\nu}%
\ell_{\rho}\ell_{\sigma}\ell_{\tau}f\left(  \ell^{2}\right)  \rightarrow
0,\ldots\label{symodd}%
\end{equation}
and to make the substitutions:%

\begin{equation}
\ell_{\mu}\ell_{\nu}f\left(  \ell^{2}\right)  \rightarrow\frac{g_{\mu\nu}}%
{n}\ell^{2}f\left(  \ell^{2}\right)  \label{symi2}%
\end{equation}
and%
\begin{equation}
\ell_{\mu}\ell_{\nu}\ell_{\rho}\ell_{\sigma}f\left(  \ell^{2}\right)
\rightarrow\frac{g_{\mu\nu}g_{\rho\sigma}+g_{\mu\rho}g_{\nu\sigma}%
+g_{\mu\sigma}g_{\rho\nu}}{n\left(  n+2\right)  }\left(  \ell^{2}\right)
^{2}f\left(  \ell^{2}\right)  , \label{symi4}%
\end{equation}
where $n$ is the space-time dimension.

\begin{center}%
\begin{figure}[ht]%
\begin{center}
\includegraphics[
height=1.5065in,
width=2.8046in
]%
{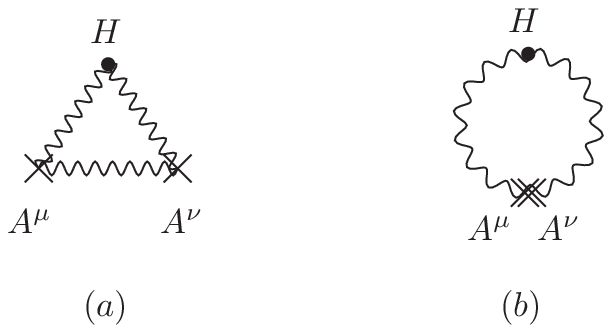}%
\end{center}
\caption{All internal lines are vector fields.}
\label{hgg-w}%
\end{figure}%

\begin{figure}[ht]%
\begin{center}
\includegraphics[
height=4.5247in,
width=5.8643in
]%
{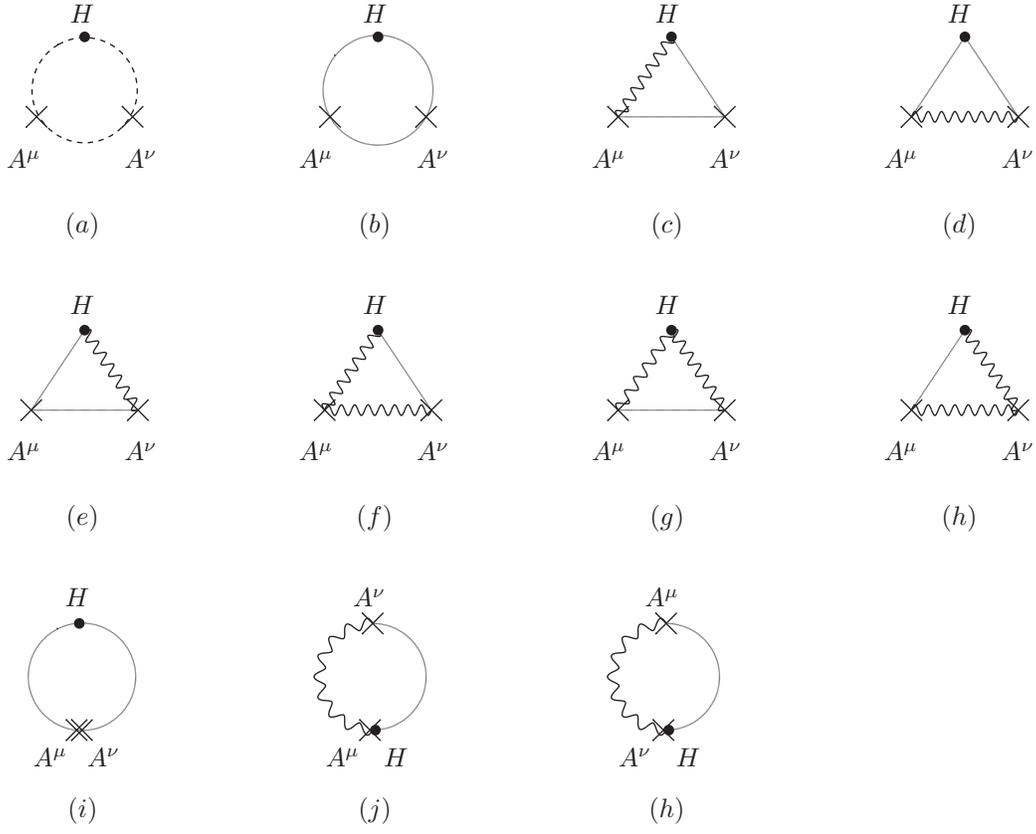}%
\end{center}
\caption{Not all internal lines are vector fields.}%
\label{hgg-fiG}%
\end{figure}%

\end{center}

The outgoing momenta are $p$ for photon field $A^{\mu}$, $q$ for photon field
$A^{\nu}$ and the incoming momentum for $H$ is $p+q$. The on-shell conditions
are $p^{2}=q^{2}=0$ and $\left(  p+q\right)  ^{2}=M_{H}^{2}=\lambda M_{W}^{2}%
$. The polarizations for both external photons are transverse and we may set
$p^{\mu}=q^{\nu}=0$. Since we may shift or reflect the loop variable and all
poles of the propagators are at $\ell^{2}=M_{W}^{2}$ for both Feynman and
unitary gauges, the resulting integrand for any Feynman diagram can be
expressed as combinations of the following four expressions $\left(
\ref{Ia}\right)  $-$\left(  \ref{Id}\right)  $.
\begin{equation}
I_{1}\equiv\frac{1}{\left(  \ell^{2}-M_{W}^{2}\right)  } \label{Ia}%
\end{equation}%
\begin{equation}
I_{11}\equiv\frac{1}{\left(  \ell^{2}-M_{W}^{2}\right)  ^{2}} \label{Ib}%
\end{equation}%
\begin{equation}
I_{13}\equiv\frac{1}{\left(  \ell^{2}-M_{W}^{2}\right)  \left(  \left(
\ell+p+q\right)  ^{2}-M_{W}^{2}\right)  } \label{Ic}%
\end{equation}%
\begin{equation}
I_{123}\equiv\frac{1}{\left(  \ell^{2}-M_{W}^{2}\right)  \left(  \left(
\ell+p\right)  ^{2}-M_{W}^{2}\right)  \left(  \left(  \ell+p+q\right)
^{2}-M_{W}^{2}\right)  } \label{Id}%
\end{equation}
The result for the unitary gauge and for the Feynman gauge are tabulated in
Table $\left(  \ref{tb-1}\right)  $ and Table $\left(  \ref{tb-2}\right)  $.

\begin{center}%
\begin{table}[ht] \centering
\begin{tabular}
[c]{|l|l|}\hline
Diagram & Unitary Gauge\\\hline
Fig. $\ref{hgg-w}\left(  a\right)  $ & $%
\begin{array}
[c]{c}%
\frac{2\left(  n-1\right)  }{nM_{W}^{4}}g_{\mu\nu}+\left(  \frac{4-n\left(
2-3\lambda+2n\lambda\right)  }{2n\left(  n-1\right)  M_{W}^{2}}g_{\mu\nu
}+\frac{\left(  n-2\right)  \left(  2+\lambda\right)  }{2\left(  n-1\right)
\lambda M_{W}^{4}}q_{\mu}p_{\nu}\right)  I_{1}\\
+\frac{\left(  16+14\lambda-3\lambda^{2}+2n\left(  -4-6\lambda+\lambda
^{2}\right)  \right)  }{4\left(  n-1\right)  }g_{\mu\nu}I_{13}-\frac{\left(
8+\left(  8-6n\right)  \lambda+\left(  n-2\right)  \lambda^{2}\right)
}{4\left(  n-1\right)  \lambda M_{W}^{2}}q_{\mu}p_{\nu}I_{13}\\
+\left(  -8\lambda M_{W}^{2}g_{\mu\nu}+16q_{\mu}p_{\nu}\right)  I_{123}%
+4\left(  -2+2n+\lambda\right)  \left(  \ell_{\mu}p_{\nu}+\ell_{\mu}\ell_{\nu
}\right)  I_{123}%
\end{array}
$\\\hline
Fig. $\ref{hgg-w}\left(  b\right)  $ & $%
\begin{array}
[c]{c}%
\frac{-2\left(  n-1\right)  }{nM_{W}^{4}}g_{\mu\nu}+\left(  \frac{-4+n\left(
2-3\lambda+2n\lambda\right)  }{2n\left(  n-1\right)  M_{W}^{2}}g_{\mu\nu
}-\frac{\left(  n-2\right)  \left(  2+\lambda\right)  }{2\left(  n-1\right)
\lambda M_{W}^{4}}q_{\mu}p_{\nu}\right)  I_{1}\\
+\left(  \frac{\left(  24+8n^{2}+10\lambda-3\lambda^{2}+2n\left(
-12-4\lambda+\lambda^{2}\right)  \right)  }{-4\left(  n-1\right)  }g_{\mu\nu
}+\frac{\left(  8+\left(  8-6n\right)  \lambda+\left(  n-2\right)  \lambda
^{2}\right)  }{4\left(  n-1\right)  \lambda M_{W}^{2}}q_{\mu}p_{\nu}\right)
I_{13}%
\end{array}
$\\\hline
SUM & $\left(  2-2n-\lambda\right)  g_{\mu\nu}I_{13}+\left(
\begin{array}
[c]{c}%
-8\lambda M_{W}^{2}g_{\mu\nu}+16q_{\mu}p_{\nu}\\
+4\left(  \lambda+2n-2\right)  \left(  \ell_{\mu}p_{\nu}+\ell_{\mu}\ell_{\nu
}\right)
\end{array}
\right)  I_{123}$\\\hline
\end{tabular}
\caption{Unitary gauge amplitudes with ($e^{2}gM_{W})$ factored out.}%
\label{tb-1}%
\end{table}%

\begin{table}[ht] \centering
\begin{tabular}
[c]{|l|l|}\hline
Diagram & Feynman Gauge\\\hline
Fig. $\ref{hgg-w}\left(  a\right)  $ & $%
\begin{array}
[c]{c}%
2g_{\mu\nu}I_{11}+2g_{\mu\nu}I_{13}\\
+\left(
\begin{array}
[c]{c}%
M_{W}^{2}\left(  4-5\lambda\right)  g_{\mu\nu}+10q_{\mu}p_{\nu}\\
+\left(  8n-14\right)  \ell_{\mu}p_{\nu}+2q_{\mu}\ell_{\nu}+\left(
8n-12\right)  \ell_{\mu}\ell_{\nu}%
\end{array}
\right)  I_{123}%
\end{array}
$\\\hline
Fig. $\ref{hgg-w}\left(  b\right)  $ & $2\left(  1-n\right)  g_{\mu\nu}I_{13}%
$\\\hline
Fig. $\ref{hgg-fiG}\left(  a\right)  $ & $-2\left(  \ell_{\mu}p_{\nu}%
+\ell_{\mu}\ell_{\nu}\right)  I_{123}$\\\hline
Fig. $\ref{hgg-fiG}\left(  b\right)  $ & $4\lambda\left(  \ell_{\mu}p_{\nu
}+\ell_{\mu}\ell_{\nu}\right)  I_{123}$\\\hline
Fig. $\ref{hgg-fiG}\left(  c\right)  $ & $2\left(  2q_{\mu}p_{\nu}+2q_{\mu
}\ell_{\nu}+\ell_{\mu}p_{\nu}+\ell_{\mu}\ell_{\nu}\right)  I_{123}$\\\hline
Fig. $\ref{hgg-fiG}\left(  d\right)  $ & $-\lambda M_{W}^{2}g_{\mu\nu}I_{123}%
$\\\hline
Fig. $\ref{hgg-fiG}\left(  e\right)  $ & $2\left(  \ell_{\mu}\ell_{\nu}%
-\ell_{\mu}p_{\nu}\right)  I_{123}$\\\hline
Fig. $\ref{hgg-fiG}\left(  f\right)  $ & $-g_{\mu\nu}I_{13}+\left(  -M_{W}%
^{2}\left(  1+\lambda\right)  g_{\mu\nu}+2q_{\mu}p_{\nu}+5\ell_{\mu}p_{\nu
}-2q_{\mu}\ell_{\nu}+\ell_{\mu}\ell_{\nu}\right)  I_{123}$\\\hline
Fig. $\ref{hgg-fiG}\left(  g\right)  $ & $-2M_{W}^{2}g_{\mu\nu}I_{123}%
$\\\hline
Fig. $\ref{hgg-fiG}\left(  h\right)  $ & $-g_{\mu\nu}I_{13}+\left(  -M_{W}%
^{2}\left(  1+\lambda\right)  g_{\mu\nu}+3\ell_{\mu}p_{\nu}-4q_{\mu}\ell_{\nu
}+\ell_{\mu}\ell_{\nu}\right)  I_{123}$\\\hline
Fig. $\ref{hgg-fiG}\left(  i\right)  $ & $-\lambda g_{\mu\nu}I_{13}$\\\hline
Fig. $\ref{hgg-fiG}\left(  j\right)  $ & $-g_{\mu\nu}I_{11}$\\\hline
Fig. $\ref{hgg-fiG}\left(  k\right)  $ & $-g_{\mu\nu}I_{11}$\\\hline
SUM & $\left(  2-2n-\lambda\right)  g_{\mu\nu}I_{13}+\left(
\begin{array}
[c]{c}%
-8\lambda M_{W}^{2}g_{\mu\nu}+16q_{\mu}p_{\nu}\\
+4\left(  \lambda+2n-2\right)  \left(  \ell_{\mu}p_{\nu}+\ell_{\mu}\ell_{\nu
}\right)
\end{array}
\right)  I_{123}$\\\hline
\end{tabular}
\caption{Feynman gauge amplitudes with ($e^{2}gM_{W})$ factored out.}%
\label{tb-2}%
\end{table}%
\end{center}
Note that dimensional regularization is utilized in the manipulation of the
Feynman integrand of every individual diagram. The last rows of Table $\left(
\ref{tb-1}\right)  $ and Table $\left(  \ref{tb-2}\right)  $ prove that the
sum of integrands for the unitary gauge is the same as that for the Feynman
gauge. Gauge invariance is not broken.

\section{Conclusion}

In the diagrammatic proof of gauge invariance for on-shell amplitude, adopting
dimensional regularization ensures the legitimacy to shift or reflect loop
momenta and hence guarantees gauge symmetry. The sum of integrands for the
$R_{\xi}$ gauge should be $\xi$ independent.

The different result obtained from Wu's \cite{TTW} treatment of unitary gauge
stems from the peculiar handling the divergent expression of the Feynman
amplitude. In fact, it is possible to extract any value we want out of
unregularized divergent expression. For example, if we make a shift of the
loop variable $\ell\rightarrow\ell+\Delta$ on the integrand $\frac{1}{\ell
^{2}-m^{2}}$ which yields quadratically divergent integral. The difference can
be written as
\begin{align*}
&  \frac{1}{\left(  \ell+\Delta\right)  ^{2}-m^{2}}-\frac{1}{\ell^{2}-m^{2}%
}=\Delta^{\mu_{1}}\Delta^{\mu_{2}}\left[  \frac{4\ell_{\mu_{1}}\ell_{\mu_{2}}%
}{\left(  \ell^{2}-m^{2}\right)  ^{3}}-\frac{g_{\mu_{1}\mu_{2}}}{\left(
\ell^{2}-m^{2}\right)  ^{2}}\right] \\
&  =\Delta^{2}\left[  \frac{4\ell^{2}}{n\left(  \ell^{2}-m^{2}\right)  ^{3}%
}-\frac{1}{\left(  \ell^{2}-m^{2}\right)  ^{2}}\right] \\
&  =\left\{
\begin{array}
[c]{c}%
0,\;\;\text{with dimensional regularization}\\
\frac{\Delta^{2}m^{2}}{\left(  \ell^{2}-m^{2}\right)  ^{3}},\;n=4\text{
}\ \text{without dimensional regularization}%
\end{array}
\right.  .
\end{align*}
As expected, the above difference due to a shift of loop momentum vanishes
with dimensional regularization. But it is a finite value proportional to
$\Delta^{2}$ with $n=4$ and without dimensional regularization. Both
expressions $(86)$ and $\left(  87\right)  $ in Wu's \cite{TTW} treatment
would have vanished using dimensional regulazarion.

The on-shell amplitude obtained in the unitary gauge should be the same as
that in the $R_{\xi}$ gauge provided the scheme of dimensional regularization
is utilized.

\appendix\appendixpage

\section{The Electroweak Interaction terms}

Define the $2\times2$\ maxtrix%
\[
T_{a}\equiv\frac{\sigma_{a}}{2},\;\;\;a\in\left\{  1,2,3\right\}
\]
where
\[
\sigma_{1}=\left[
\begin{array}
[c]{cc}%
0 & 1\\
1 & 0
\end{array}
\right]  ,\sigma_{2}=\left[
\begin{array}
[c]{cc}%
0 & -i\\
i & 0
\end{array}
\right]  ,\sigma_{3}=\left[
\begin{array}
[c]{cc}%
1 & 0\\
0 & -1
\end{array}
\right]
\]
are the Pauli matrices. Define the maxtrix $W$ vector fields and ghost fields by%

\[
W\equiv\sum_{a=1}^{3}W_{a}T_{a},\bar{c}\equiv\sum_{a=1}^{3}\bar{c}_{a}%
T_{a},c\equiv\sum_{a=1}^{3}c_{a}T_{a},
\]
Write the $2\times1$ complex scalar field $\phi$ as
\[
\phi=\frac{1}{\sqrt{2}}\left[
\begin{array}
[c]{c}%
i\phi_{1}+\phi_{2}\\
H+\upsilon-i\phi_{3}%
\end{array}
\right]
\]
where the component fields $H$, $\phi_{1}$, $\phi_{2}$, $\phi_{3}$ are all
real scalar fields. The electroweak Lagrangian is%
\begin{equation}
L=-\frac{1}{4}F_{B}^{\mu\nu}F_{B\mu\nu}-\frac{1}{2}Tr\left(  F_{W}^{\mu\nu
}F_{W\mu\nu}\right)  +\left(  D_{\mu}\phi\right)  ^{\dagger}\left(  D^{\mu
}\phi\right)  -\frac{\lambda}{8}g_{W}^{2}\left(  \phi^{\dagger}\phi
-\frac{\upsilon^{2}}{2}\right)  ^{2} \label{L1}%
\end{equation}
where
\[
F_{W}^{\mu\nu}=\partial^{\mu}W^{\nu}-\partial^{\nu}W^{\mu}+ig_{W}\left[
W^{\mu},W^{\nu}\right]
\]%
\[
F_{B}^{\mu\nu}=\partial^{\mu}B^{\nu}-\partial^{\nu}B^{\mu}%
\]
and%
\[
D_{\mu}\phi\equiv\partial_{\mu}\phi+i\left(  g_{W}W_{\mu}-\frac{1}{2}%
g_{B}B_{\mu}\right)  \phi
\]
The electroweak Lagrangian $\left(  \ref{L1}\right)  $ is invariant under the
$BRST$ \cite{BRS}\ variations
\begin{subequations}
\label{0}%
\begin{equation}
\delta_{B}\phi=i\left(  \frac{g_{B}}{2}c_{B}-g_{W}c\right)  \phi\label{brsfi}%
\end{equation}%
\end{subequations}
\begin{equation}
\delta_{B}W_{\mu}=\partial_{\mu}c+ig_{W}\left[  W_{\mu},c\right]  \label{brsw}%
\end{equation}%
\begin{equation}
\delta_{B}B_{\mu}=\partial_{\mu}c_{B} \label{brsb}%
\end{equation}
Using $\left(  \ref{brsfi}\right)  $, $\left(  \ref{brsw}\right)  $ and%
\[
2Tr\left(  T_{a}T_{b}\right)  =\delta_{ab},\;\left[  T_{a},T_{b}\right]
=\sum_{c}if_{abc}T_{c},
\]
we may derive the BRST variabtions for the component fields:
\[
\delta_{B}W_{a\mu}=2Tr\left(  \delta_{B}W_{\mu}T_{a}\right)  =\partial_{\mu
}c_{a}+g_{W}f_{abc}c_{b}W_{c\mu}%
\]%
\[
\delta_{B}H=\frac{1}{2}\left(  g_{B}c_{B}\phi_{3}+g_{W}\sum c_{a}\phi
_{a}\right)
\]%
\[
\delta_{B}\phi_{3}=-\frac{1}{2}\left(  g_{B}c_{B}+g_{W}c_{3}\right)  \left(
H+\upsilon\right)  +\frac{1}{2}g_{W}\left(  c_{1}\phi_{2}-c_{2}\phi
_{1}\right)
\]%
\[
\delta_{B}\phi_{1}=-\frac{1}{2}g_{W}c_{1}\left(  H+\upsilon\right)  -\frac
{1}{2}g_{B}c_{B}\phi_{2}+\frac{1}{2}g_{W}\left(  c_{2}\phi_{3}-c_{3}\phi
_{1}\right)
\]%
\[
\delta_{B}\phi_{2}=-\frac{1}{2}g_{W}c_{2}\left(  H+\upsilon\right)  -\frac
{1}{2}g_{B}c_{B}\phi_{1}+\frac{1}{2}g_{W}\left(  c_{3}\phi_{1}-c_{1}\phi
_{3}\right)
\]
Define the fields $A$ and $Z$ fields in terms of $B\,\ $and $W_{3}$,%
\[
\left[
\begin{array}
[c]{c}%
A_{\mu}\\
Z_{\mu}%
\end{array}
\right]  \equiv\frac{1}{M_{Z}}\left[
\begin{array}
[c]{cc}%
M_{W} & -M_{B}\\
M_{B} & M_{W}%
\end{array}
\right]  \left[
\begin{array}
[c]{c}%
B_{\mu}\\
W_{3\mu}%
\end{array}
\right]
\]
and similarly for the corresponding ghost and antighost fields:%
\[
\left[
\begin{array}
[c]{c}%
c_{A}\\
c_{Z}%
\end{array}
\right]  \equiv\frac{1}{M_{Z}}\left[
\begin{array}
[c]{cc}%
M_{W} & -M_{B}\\
M_{B} & M_{W}%
\end{array}
\right]  \left[
\begin{array}
[c]{c}%
c_{B}\\
c_{3}%
\end{array}
\right]
\]%
\[
\left[
\begin{array}
[c]{c}%
\bar{c}_{A}\\
\bar{c}_{Z}%
\end{array}
\right]  \equiv\frac{1}{M_{Z}}\left[
\begin{array}
[c]{cc}%
M_{W} & -M_{B}\\
M_{B} & M_{W}%
\end{array}
\right]  \left[
\begin{array}
[c]{c}%
\bar{c}_{B}\\
\bar{c}_{3}%
\end{array}
\right]
\]
where%
\[
M_{B}=\frac{1}{2}g_{B}\upsilon,M_{W}=\frac{1}{2}g_{W}\upsilon
\]
and%
\[
M_{Z}=\sqrt{M_{B}^{2}+M_{W}^{2}}.
\]
The square of mass $M_{H}$ for the physical Higgs $H$ field is%
\[
M_{H}^{2}=\lambda M_{W}^{2}\text{.}%
\]
The electric charge $e$ is%
\[
e=\frac{g_{B}g_{W}}{\sqrt{g_{B}^{2}+g_{W}^{2}}}.
\]
Adding the gauge fixing terms%
\[
L_{gf}=-\frac{1}{2}\left(  \partial_{\mu}A^{\mu}\right)  ^{2}-\frac{1}{2\xi
}\left[  \sum_{a=1,2}\left(  \partial_{\mu}W_{a}^{\mu}-\xi M_{W}\phi
_{a}\right)  ^{2}+\left(  \partial_{\mu}Z^{\mu}-\xi M_{Z}\phi_{3}\right)
^{2}\right]
\]
and ghost terms%
\[
L_{gh}=i\bar{c}_{A}\delta_{B}\left(  \partial_{\mu}A^{\mu}\right)  +i\bar
{c}_{Z}\delta_{B}\left(  \partial_{\mu}Z^{\mu}-\xi M_{Z}\phi_{3}\right)
+\sum_{a=1,2}i\bar{c}_{a}\delta_{B}\left(  \partial_{\mu}W_{a}^{\mu}-\xi
M_{W}\phi_{a}\right)
\]
The effective Lagrangian \cite{CTBRS} for the electroweak theory is%
\begin{equation}
L_{eff}=L+L_{gf}+L_{gh} \label{Leff}%
\end{equation}

\section{The Free Propagators}

The terms quadratic in field variables in the effective Lagrangian $\left(
\ref{Leff}\right)  $ lead to the free propagators for fields $W_{1}$, $W_{2}$,
$A$, $Z$, $H$, $\phi_{1}$, $\phi_{2}$, $\phi_{3}$ and ghost fields $\bar
{c}_{1},c_{1},\bar{c}_{2},c_{2},\bar{c}_{A},c_{A},\bar{c}_{Z},c_{Z}$ in the
$R_{\xi}$ gauge:%
\[
D\left(  W_{a\mu},W_{b\nu};k\right)  =\delta_{ab}\frac{-i}{k^{2}-M_{W}^{2}%
}\left(  g^{\mu\nu}-\frac{k^{\mu}k^{\mu}}{k^{2}}\right)  +\delta_{ab}\frac
{-i}{\frac{k^{2}}{\xi}-M_{W}^{2}}\frac{k^{\mu}k^{\mu}}{k^{2}}\;\;a,b\in
\left\{  1,2\right\}
\]%
\[
D\left(  A_{\mu},A_{\nu};k\right)  =\frac{-ig^{\mu\nu}}{k^{2}},\;D\left(
Z_{\mu},Z_{\nu};k\right)  =\frac{-i}{k^{2}-M_{z}^{2}}\left(  g^{\mu\nu}%
-\frac{k^{\mu}k^{\mu}}{k^{2}}\right)  +\frac{-i}{\frac{k^{2}}{\xi}-M_{z}^{2}%
}\frac{k^{\mu}k^{\mu}}{k^{2}}%
\]%
\[
D\left(  \phi_{a},\phi_{b};k\right)  =\frac{i}{k^{2}-\xi M_{W}^{2}}%
\;a,b\in\left\{  1,2\right\}
\]%
\[
D\left(  \phi_{3},\phi_{3};k\right)  =\frac{i}{k^{2}-\xi M_{Z}^{2}},\;D\left(
H,H;k\right)  =\frac{i}{k^{2}-M_{H}^{2}}%
\]%
\[
D\left(  \bar{c}_{a},c_{b};k\right)  =\delta_{ab}\frac{1}{k^{2}-\xi M_{W}^{2}%
},\;a,b\in\left\{  1,2\right\}
\]%
\[
D\left(  \bar{c}_{A},c_{A};k\right)  =\frac{1}{k^{2}},\;D\left(  \bar{c}%
_{Z},c_{Z};k\right)  =\frac{1}{k^{2}-\xi M_{Z}^{2}}%
\]
The transverse parts of vector fields, $W_{1}$, $W_{2}$, $A$, $Z$ and $H$ are
physical fields as their propagators are independent of the gauge parameter
$\xi$. \ The propagators for the unitary gauge are the limit $\xi
\rightarrow\infty$ of the above propagators in the $R_{\xi}$ gauge:%
\[
D\left(  W_{a\mu},W_{b\nu};k\right)  =\delta_{ab}\frac{-i}{k^{2}-M_{W}^{2}%
}\left(  g^{\mu\nu}-\frac{k^{\mu}k^{\mu}}{M_{W}^{2}}\right)  \;\;a,b\in
\left\{  1,2\right\}
\]%
\[
D\left(  A_{\mu},A_{\nu};k\right)  =\frac{-ig^{\mu\nu}}{k^{2}},\;D\left(
Z_{\mu},Z_{\nu};k\right)  =\frac{-i}{k^{2}-M_{z}^{2}}\left(  g^{\mu\nu}%
-\frac{k^{\mu}k^{\mu}}{M_{z}^{2}}\right)
\]%
\[
D\left(  \phi_{a},\phi_{b};k\right)  =0\;a,b\in\left\{  1,2\right\}
\]%
\[
D\left(  \phi_{3},\phi_{3};k\right)  =0,\;D\left(  H,H;k\right)  =\frac
{i}{k^{2}-M_{H}^{2}}%
\]%
\[
D\left(  \bar{c}_{a},c_{b};k\right)  =0,\;a,b\in\left\{  1,2\right\}
\]%
\[
D\left(  \bar{c}_{A},c_{A};k\right)  =\frac{1}{k^{2}},\;D\left(  \bar{c}%
_{Z},c_{Z};k\right)  =0
\]
In the unitary gauge, all the propagators for the non-physical $\phi_{1}$,
$\phi_{2}$, $\phi_{3}$ fields and ghost fields $\bar{c}_{1},c_{1},\bar{c}%
_{2},c_{2},\bar{c}_{Z},c_{Z}$ vanish.

\end{document}